\documentclass[twocolumn,table]{aastex631}
\usepackage{xspace}
\usepackage{amsmath}
\let\tablenum\relax
\usepackage{siunitx}

\newcommand\Lya{Ly$\alpha$\xspace}
\newcommand{\sbunits}{\mathrm{erg}\,\mathrm{s}^{-1}\,\mathrm{cm}^{-2}\,\mathrm{arcsec}^{-2}\,{\rm \AA}^{-1}}

\newcommand{\lowzmonosignificance}[0]{$\simeq 8\sigma$\xspace}

\begin{document}

\title{SIMPLE: Simple Intensity Map Producer for Line Emission}

\author[0000-0002-6907-8370]{Maja Lujan Niemeyer}
\affiliation{Max-Planck-Institut f\"{u}r Astrophysik, Karl-Schwarzschild-Str. 1, 85741 Garching, Germany}
\email{maja@mpa-garching.mpg.de}

\author[0000-0002-0961-4653]{Jos{\'e} Luis Bernal}
\affiliation{Instituto de Física de Cantabria (IFCA), CSIC-Univ. de Cantabria, Avda. de los Castros s/n, E-39005 Santander, Spain}
\affiliation{Max-Planck-Institut f\"{u}r Astrophysik, Karl-Schwarzschild-Str. 1, 85741 Garching, Germany}

\author[0000-0002-0136-2404]{Eiichiro Komatsu}
\affiliation{Max-Planck-Institut f\"{u}r Astrophysik, Karl-Schwarzschild-Str. 1, 85741 Garching, Germany}
\affiliation{Kavli Institute for the Physics and Mathematics of the Universe (Kavli IPMU, WPI), University of Tokyo, Chiba 277-8582, Japan}

\begin{abstract}

We present the Simple Intensity Map Producer for Line Emission (\textsc{Simple}), a public code for quickly simulating mock line-intensity maps, and an analytical framework for modeling intensity maps including observational effects. \textsc{Simple} can be applied to any spectral line sourced by galaxies.
The \textsc{Simple} code is based on lognormal mock catalogs of galaxies including positions and velocities and assigns luminosities following the luminosity function. After applying a selection function to distinguish between detected and undetected galaxies, the code generates an intensity map, which can be modified with anisotropic smoothing, noise, a mask, and sky subtraction, and calculates the power spectrum multipoles.
We show that the intensity autopower spectrum and the galaxy-intensity cross-power spectrum agree well with the analytical estimates in real space.
We derive and show that the sky subtraction suppresses the intensity autopower spectrum and the cross-power spectrum on scales larger than the size of an individual observation.
As an example application, we make forecasts for the sensitivity of an intensity mapping experiment similar to the Hobby-Eberly Telescope Dark Energy Experiment (HETDEX) to the cross-power spectrum of Ly$\alpha$-emitting galaxies and the Ly$\alpha$ intensity. We predict that HETDEX will detect the galaxy-intensity cross-power spectrum on scales of $0.04\, h\,\mathrm{Mpc}^{-1} < k < 1\, h\,\mathrm{Mpc}^{-1}$.
\end{abstract}


\section{Introduction} \label{sec:intro}

Line-intensity mapping (LIM) is a promising survey strategy for constraining cosmological parameters
and studying astrophysics of galaxies and intergalactic gas \citep[e.g.,][]{kovetz/etal:2017,bernal/kovetz:2022}. 
Instead of relying on detections of individual galaxies to trace the large-scale structure of matter, 
LIM measures the integrated line emission from all galaxies and the intergalactic medium (IGM) as a biased tracer of the matter distribution.
It collects light from all emitters, including those that are too faint for individual detection at high redshift,
and requires lower resolution and shorter integration times than traditional galaxy surveys.

Several LIM surveys are in operation \citep{santos/etal:2016,keating/etal:2016,keating/etal:2020,deboer/etal:2017,concerto/etal:2020,gebhardt/etal:2021,cleary/etal:2022}
or in preparation \citep{dore/etal:2014,vieira/etal:2020,sun/etal:2021,switzer/etal:2021,ccat_prime/etal:2023}. They target lines ranging from the $21\,\mathrm{cm}$ line in radio frequencies to the ultraviolet Lyman-$\alpha$ (\Lya) line, which trace atomic or molecular gas. Although the $21\,\mathrm{cm}$ line is emitted by neutral atomic hydrogen in the IGM and in neutral pockets within the interstellar medium (ISM) of galaxies, other lines targeted by LIM experiments are associated with star formation in the ISM \citep[e.g.,][]{bernal/kovetz:2022}. Neglecting scattering and diffuse emission of the \Lya line, which can illuminate the circumgalactic medium (CGM) and IGM \citep[see, e.g.,][]{byrohl/etal:2021,byrohl/nelson:2023}, these line intensities are therefore sourced within galaxies and influenced by their astrophysical properties such as their star-formation rate.

Mock intensity maps are necessary to model the signal beyond the capabilities of analytical models, estimate statistical covariance,
and explore observational effects such as foregrounds, sky subtraction, and survey footprints on summary statistics \citep[e.g.,][]{cunnington/etal:2023b}.
Modeling techniques of line-intensity maps have to compromise between astrophysical complexity, volume, and computational feasibility. 
On the one hand, cosmological hydrodynamical simulations can be post-processed 
using the astrophysical properties of galaxies and the gas to infer the line emission \citep[e.g.,][]{moriwaki/etal:2019,silva/etal:2021,kannan/etal:2022,byrohl/nelson:2023,liang/etal:2023}. Halo catalogs from N-body simulations can also be combined with galaxy evolution models to predict line luminosities and produce intensity maps \citep[e.g.,][]{lidz/etal:2011,gong/etal:2012,li/etal:2016,chung/etal:2019,spina/porciani/schimd:2021,bethermin/etal:2022,moradinezhad/etal:2022b,sato-polito/kokron/bernal:2022}.
Although these approaches provide accurate small-scale clustering and include astrophysical dependences of the line intensity, 
it is not feasible to produce enough realizations for covariance estimation or parameter inference.
To speed up the calculation, one can model the underlying dark matter density field through Lagrangian perturbation theory or mass-peak patch and apply various post-processing steps to model the astrophysical dependence of line intensities \citep[][]{mesinger/furlanetto/cen:2011,silva/etal:2013,silva/etal:2015,mesinger/greig/sobacchi:2016,heneka/cooray/feng:2017,heneka/mesinger:2020,masribas/etal:2023,chung/etal:2022,roy/etal:2023}{}{}.
The line intensity can also be modeled by multiplying the total matter density in a fast lognormal simulation by a bias \citep{alonso/ferreira/santos:2014,rubiola/cunnington/camera:2022}.
These approaches are fast and enable simulating many realizations of large volumes, but do not account for the shot-noise contribution to the power spectrum from the discreteness of the line-emitting sources.
Finally, \citet{obuljen/etal:2022} use a field-level forward-modeling approach to simulate intensity maps based on effective field theory.

To include shot noise, one can generate a galaxy catalog from a lognormal galaxy number density field via Poisson sampling.
Lognormal simulations take advantage of the roughly lognormal probability density function (PDF) of matter and galaxy density distributions,
 measured both in N-body simulations \citep[e.g.,][]{kayo/taruya/suto:2001,shin/etal:2017} and in galaxy surveys \citep[e.g.,][]{clerkin/etal:2017}. 
 The lognormality of the PDF of the density contrast $\delta$ implies that the logarithmic transformation field, $\mathrm{ln}\left(1+\delta\right)$, is a Gaussian random field whose statistics are defined entirely by its two-point correlation function or power spectrum.
\citet{agrawal/etal:2017} present a public lognormal mock generator for galaxy catalogs, including self-consistent velocities of the galaxies.
The velocities enable redshift-space distortion (RSD) modeling beyond the linear Kaiser model \citep{kaiser:1987}, 
which is critical for any large-scale structure measurement in redshift space such as LIM.

The lognormal galaxy catalog generator of \citet{agrawal/etal:2017} has been extended to generate weak-lensing fields \citep[][]{makiya/kayo/komatsu:2021}{}{}. 
In this paper, we extend it to quickly and self-consistently generate intensity maps 
and galaxy catalogs, which we call the Simple Intensity Map Producer for Line Emission (\textsc{Simple}).\footnote{\url{https://github.com/mlujnie/simple} \citep[][]{simple_code}{}{}}
Given a luminosity function for any emission line, we assign line luminosities to galaxies in a lognormal galaxy simulation, 
apply a selection function to obtain a galaxy catalog, and calculate the intensity on a grid.
We can add noise, smooth the intensity map, model the sky subtraction, and apply a mask before calculating the galaxy and LIM auto- and cross-power spectra.
The simplicity of this approach enables us to quickly generate many mock intensity maps, e.g., to estimate the covariance matrices of the LIM power spectra.

As an example of its capabilities, we use the \textsc{Simple} framework to forecast the sensitivity of the Hobby-Eberly Telescope Dark Energy Experiment \citep[HETDEX;][]{gebhardt/etal:2021}
to the cross correlation of Ly$\alpha$-emitting galaxies (LAEs) and the \Lya intensity. 
An analytical forecast for the intensity autopower spectrum in HETDEX was presented in
\citet[][]{fonseca/etal:2017}.

This paper is structured as follows. 
Section \ref{sec:power_spectrum_modeling} derives the power spectrum formalism.
In Section \ref{sec:simple_description} we describe the steps of the \textsc{Simple} code
to generate line-intensity mocks. 
Section \ref{sec:example_mock_setup} explains an example mock setup for HETDEX that is used in the rest of the paper.
We validate the \textsc{Simple} code by comparing the power spectrum multipoles with theoretical predictions in Section \ref{sec:validation}.
Section \ref{sec:hetdex_forecast} presents the forecast for the LAE-\Lya intensity cross-correlation of HETDEX.
Section \ref{sec:discussion} discusses the limitations of the \textsc{Simple} framework.
We conclude in Section \ref{sec:summary}.

 We use the following Fourier convention:
 \begin{equation}
 \begin{split}
     \tilde{f}(\mathbf{k}) &= \int \mathrm{d}^3\mathbf{x} f(\mathbf{x}) e^{i \mathbf{k} \cdot \mathbf{x}}\,, \\
     f(\mathbf{x}) &= \int \frac{\mathrm{d}^3 \mathbf{k}}{(2\pi)^3} \tilde{f}(\mathbf{k})e^{-i\mathbf{k} \cdot \mathbf{x}}\,,
\end{split}
\end{equation}
where the tilde denotes quantities in Fourier space. 
We refer to \emph{real} space in contrast to {redshift} space,
and to \emph{configuration} space in contrast to {Fourier} space.

Throughout this paper, we assume a flat $\Lambda$ cold dark matter ($\Lambda$\textsc{CDM}) cosmology with $H_0 = 67.66\,\mathrm{km\,s^{-1}\,Mpc^{-1}}$, $\Omega_{\mathrm{b},0}h^2 = 0.022$, $\Omega_{\mathrm{m},0}h^2=0.142$, $\ln\left(10^{10} A_s\right)=3.094$, and $n_s = 0.9645$.

\section{Power spectrum modeling}
\label{sec:power_spectrum_modeling}

Consider a fluctuation $\delta A(\mathbf{x}) = A(\mathbf{x}) - \langle A(\mathbf{x})\rangle$ of a field $A$, such as the intensity $A(\mathbf{x}) = I(\mathbf{x})$ and the normalized galaxy number density
$A(\mathbf{x}) = n(\mathbf{x}) / \langle n(\mathbf{x}) \rangle = 1 + \delta_\mathrm{g}(\mathbf{x})$.
Here, $\langle \cdot (\mathbf{x}) \rangle$ denotes the mean field at location $\mathbf{x}$ over many realizations, for example, the mean intensity as a function of redshift, or the galaxy selection function as a function of position.
The dimensionless correlation function of fields $A$ and $B$ in configuration space is defined as 
\begin{equation}
\begin{split}
    \xi_{AB}(\mathbf{x} - \mathbf{y}) &= \frac{\langle A(\mathbf{x}) B(\mathbf{y}) \rangle - \langle A(\mathbf{x})  \rangle \langle B(\mathbf{y}) \rangle}{\langle A(\mathbf{x})  \rangle \langle B(\mathbf{y}) \rangle}\\
    &= \frac{\langle \delta A(\mathbf{x}) \delta B(\mathbf{y}) \rangle}{\langle A(\mathbf{x})  \rangle \langle B(\mathbf{y}) \rangle}\,.
\end{split}
\end{equation}
The corresponding power spectrum with the dimension of volume is the Fourier transform of the correlation function,
\begin{equation}
\label{eq:def_P_AB}
    P_{AB}(\mathbf{k}) = \int \mathrm{d}^3\mathbf{s}\, \xi^{AB}(\mathbf{s}) e^{i\mathbf{k}\cdot\mathbf{s}}.
\end{equation}

\subsection{Galaxy and Intensity Auto- and Cross-power Spectra}
\label{subsec:pk}

Following the standard approach of \citet[][]{peebles:1980}, we model the galaxy number $N(\mathbf{x})$ as a Poisson point process in infinitesimal volume elements $\delta V$, so that the occupation number in each cell is $N_i \in \{0,1\}$.
The expectation value of $N$ in one cell is $\Bar{n}\left[1+\delta_\mathrm{g}(\mathbf{x})\right] \delta V$, where $\Bar{n}$ is the mean number density of galaxies in the entire volume, and $\delta_\mathrm{g}$ is the galaxy overdensity.
Let each galaxy have a line luminosity $L_i$, 
which is sampled from a luminosity function ${\rm d}n/{\rm d}L$. 
We require that the integral $\int \mathrm{d}L\,\frac{\mathrm{d}n}{\mathrm{d}L} =: \Bar{n}$ converge
so that $\phi(L) := \frac{\mathrm{d}n}{\mathrm{d}L}\Bar{n}^{-1}$ is a PDF, for example, by setting a minimum luminosity. 

The specific intensity in a cell with luminosity $L(\mathbf{x})$ is 
\begin{align}
    I_\lambda(\mathbf{x}) &= \frac{c}{4 \pi \lambda_0  H(z) (1+z)^2 } \frac{L(\mathbf{x})}{\delta V} = X^\lambda_I(\mathbf{x}) \frac{L(\mathbf{x})}{\delta V},\\
    \mathrm{or} ~ I_\nu(\mathbf{x}) &= \frac{c}{4 \pi \nu_0  H(z)} \frac{L(\mathbf{x})}{\delta V} = X^\nu_I(\mathbf{x}) \frac{L(\mathbf{x})}{\delta V},
\end{align}
where $c$ is the speed of light, $H(z)$ is the Hubble expansion rate, $X_I$ is a redshift-dependent conversion factor, and $\lambda_0$ and $\nu_0$ are the rest-frame wavelength and frequency of the line, respectively. 
For simplicity, we refer to the specific intensity as `intensity' with the symbol $I$.

Given a function $f(A)$ of a continuous random variable $A$ with PDF $\phi(A)$, we can calculate its expectation value as $\langle f(A) \rangle = \int \mathrm{d}A\,f(A)\phi(A)$. 
As the PDF of $I\delta V$ at position $\mathbf{x}$ is given by $\phi^\prime(I\delta V) = \phi(L) X_I(\mathbf{x})$,
the first and second moments of $I \delta V$ are given by
\begin{align}
    \begin{split}
    \langle I(\mathbf{x}) \delta V \rangle &= \bar{N} X_I(\mathbf{x}) \int \mathrm{d}L\, \phi(L) L\\
    &= \delta V X_I(\mathbf{x}) \int \mathrm{d}L \, \frac{\mathrm{d}n}{\mathrm{d}L} L ,
    \end{split}\\
    \begin{split}
    \langle I^2(\mathbf{x}) \delta V^2 \rangle &= \Bar{N} X_I^2(\mathbf{x}) \int \mathrm{d}L\, \phi(L) L^2\\
    &= \delta V X_I^2(\mathbf{x}) \int \mathrm{d}L\, \frac{\mathrm{d}n}{\mathrm{d}L} L^2,
    \end{split}
\end{align}
where $\bar N=\bar n\delta V$.

Following the integration approach of \citet[][]{feldman/kaiser/peacock:1994}, we find 
\begin{equation}
\begin{split}
    \langle I(\mathbf{x}) I(\mathbf{y}) \rangle &= \langle I(\mathbf{x}) \rangle \langle I(\mathbf{y})\rangle \left[1+\xi_{II}(\mathbf{x} - \mathbf{y})\right] \\
    &+ \delta_\mathrm{D}(\mathbf{x}-\mathbf{y}) X_I^2(\mathbf{x}) \int \mathrm{d}L \frac{\mathrm{d}n}{\mathrm{d}L}L^2\,,
\end{split}
\end{equation}
where $\delta_{\rm D}$ is the Dirac delta function. 
The second term in the previous expression assumes Poisson shot noise.
Similarly, the cross correlation with the galaxy density contrast of detected galaxies, $\delta_\mathrm{g}(\mathbf{x}) = n(\mathbf{x})/\langle n(\mathbf{x})\rangle - 1$, is given by
\begin{equation}
\begin{split}
    \langle I(\mathbf{x}) \delta_\mathrm{g}(\mathbf{y}) \rangle &= \langle I(\mathbf{x})\rangle \left[1 + \xi_{I{\rm g}}(\mathbf{x}-\mathbf{y}) \right] \\
    &+ \delta_\mathrm{D}(\mathbf{x}-\mathbf{y}) \left[ \frac{X_I(\mathbf{x})}{\langle n(\mathbf{x}) \rangle} \int \mathrm{d}L \frac{\mathrm{d}n}{\mathrm{d}L} L \right]_{\mathcal{G_\mathrm{g}} \cap \mathcal{G_I}}\,,
\end{split}
\end{equation}
where the (second) shot-noise term only contains galaxies that contribute to the galaxy catalog and the intensity map, 
i.e., only detected galaxies with nonzero luminosity in the target line, denoted in the expression as $\mathcal{G_\mathrm{g}} \cap \mathcal{G_I}$. 

We define 
the weighted Fourier transform as $\widetilde{\delta I}(\mathbf{k}) = \int \mathrm{d}^3\mathbf{x}\, w_I(\mathbf{x}) \delta I(\mathbf{x}) e^{i\mathbf{k}\cdot\mathbf{x}}$ with a dimensionless weight $w_I(\mathbf{x})$, 
which can represent a survey footprint.\footnote{\citet{blake:2019} follow a similar approach, but use weights in units of (intensity)$^{-1}$.}  
We define the estimator for the intensity power spectrum,
\begin{equation}
    \hat{P}_{II}(\mathbf{k}) = V_{\rm box}^{-1}\langle |\widetilde{\delta I}(\mathbf{k})|^2 \rangle\,,
\end{equation}
where $V_{\rm box}$ is the volume of a cuboid enclosing the survey used to compute the Fourier transform, so that $\hat{P}_{II}(\mathbf{k})$ has the dimension of volume times intensity squared.
We find that
\begin{equation}
\label{eq:delta_I_k_sq}
\begin{split}
    \langle |\widetilde{\delta I}(\mathbf{k})|^2 \rangle &= \int \frac{\mathrm{d}^3\mathbf{k}^\prime}{(2\pi)^3} P_{II}(\mathbf{k}^\prime) |\widetilde{W}_I(\mathbf{k}-\mathbf{k}^\prime)|^2 \\
    &+ \int \mathrm{d}^3\mathbf{x}\, w_I^2(\mathbf{x}) X_I^2(\mathbf{x}) \int \mathrm{d} L \frac{\mathrm{d}n}{\mathrm{d}L} L^2,
\end{split}
\end{equation}
where $P_{II}$ is the power spectrum defined in Eq.~\eqref{eq:def_P_AB}, and
the window function is defined as $\widetilde{W}_I(\mathbf{k}) = \int \mathrm{d}^3 \mathbf{x}\, e^{i\mathbf{k}\cdot\mathbf{x}} w_I(\mathbf{x}) \langle I(\mathbf{x})\rangle$.

Similary, we define the estimator for the cross-power spectrum as
\begin{equation}
    \hat{P}_{{\rm g}I}(\mathbf{k}) = V_{\rm box}^{-1}\langle \widetilde{\delta I}(\mathbf{k}) \widetilde{\delta}_\mathrm{g}^\ast(\mathbf{k}) \rangle\,,
\end{equation}
with the dimension of volume times intensity, where the asterisk is the complex conjugate operator, and
\begin{equation}
\begin{split}
    \langle \widetilde{\delta I}(\mathbf{k}) \tilde{\delta}_\mathrm{g}^\ast(\mathbf{k}) \rangle = \int \frac{\mathrm{d}^3\mathbf{k}^\prime}{(2\pi)^3} P_{I\mathrm{g}}(\mathbf{k}^\prime) \widetilde{W}_I(\mathbf{k}-\mathbf{k}^\prime) \widetilde{W}^\ast_\mathrm{g}(\mathbf{k}-\mathbf{k}^\prime)\\
    \vspace{1cm} + \int \mathrm{d}^3\mathbf{x}\, w_I(\mathbf{x}) w_\mathrm{g}(\mathbf{x}) \left[\frac{X_I(\mathbf{x})}{\langle n(\mathbf{x}) \rangle} \int \mathrm{d} L \frac{\mathrm{d}n}{\mathrm{d}L} L\right]_{\mathcal{G_\mathrm{g}} \cap \mathcal{G_I}}\,,
\end{split}
\end{equation}
where the galaxy window function is $\widetilde{W}_\mathrm{g}(\mathbf{k}) = \int \mathrm{d}^3 \mathbf{x}\, e^{i\mathbf{k}\cdot\mathbf{x}} w_\mathrm{g}(\mathbf{x})$ with a dimensionless weight, $w_\mathrm{g}(\mathbf{x})$.

Finally, we define the galaxy power spectrum estimator as
\begin{equation}
    \hat{P}_{{\rm gg}}(\mathbf{k}) = V_{\rm box}^{-1}\langle |\tilde{\delta}_{\rm g}(\mathbf{k})|^2 \rangle\,,
\end{equation}
where
\begin{equation}
\begin{split}
    \langle |\tilde{\delta}_{\rm g}(\mathbf{k})|^2 \rangle &= \int \frac{\mathrm{d}^3\mathbf{k}^\prime}{(2\pi)^3} P_{{\rm gg}}(\mathbf{k}^\prime) |\widetilde{W}_{\rm g}(\mathbf{k}-\mathbf{k}^\prime)|^2 \\
    &+ \int \mathrm{d}^3\mathbf{x}\, \frac{w^2_\mathrm{g}(\mathbf{x})}{\langle n(\mathbf{x}) \rangle} \,.
\end{split}
\end{equation}

\subsection{Smoothing and Noise}
\label{subsec:noise}

Limited observational resolution can be modeled by smoothing the intensity map.
Suppose that the intensity is smoothed with a smoothing kernel $D(\mathbf{x})$, i.e., $\tilde{I}_\mathrm{s}(\mathbf{k}) = \tilde{I}(\mathbf{k}) \tilde{D}(\mathbf{k})$. 
Then the factors in the power spectrum estimators change as
\begin{align}
    \langle |\widetilde{\delta I}_\mathrm{s}(\mathbf{k})|^2 \rangle &= \langle |\widetilde{\delta I}(\mathbf{k})|^2 \rangle |\tilde{D}(\mathbf{k})|^2\,,\\
    \langle \widetilde{\delta I}_\mathrm{s}(\mathbf{k}) \tilde{\delta}_\mathrm{g}^\ast(\mathbf{k}) \rangle &= \langle \widetilde{\delta I}(\mathbf{k}) \tilde{\delta}_\mathrm{g}^\ast(\mathbf{k}) \rangle \tilde{D}(\mathbf{k})\,.
\end{align}
Examples for smoothing kernels are a Gaussian or top-hat smoothing in the line-of-sight (LOS) direction, mimicking the line-spread function or binning of intensity into frequency channels,
\begin{align}
    D_\parallel^\mathrm{Gauss} &= \exp\left(-\frac{1}{2} k_\parallel^2 s_\parallel^2\right),\\
    D_\parallel^\mathrm{top-hat} &= \mathrm{sinc}\left(\frac{1}{2} k_\parallel s_\parallel \right),
\end{align}
and a Gaussian smoothing in the angular direction, mimicking the beam smoothing or point-spread function (PSF), 
\begin{equation}
    D_\perp^\mathrm{Gauss} = \exp\left(-\frac{1}{2} k_\perp^2 s_\perp^2\right),
\end{equation}
where $s_\parallel$ and $s_\perp$ define the smoothing lengths parallel and perpendicular to the LOS, respectively. Similarly, $k_\parallel = \mu k$ and $k_\perp = \sqrt{1-\mu^2}k$ are the components of the wavenumber parallel and perpendicular to the LOS, respectively, and $\mu$ is the cosine of the angle between the LOS and the wavevector $\mathbf{k}$.

Uncorrelated noise $\Delta I_{\rm n}$ with zero mean can be added to the intensity to model instrumental and sky noise. Therefore, $I_\mathrm{s,n}(\mathbf{x}) = I_\mathrm{s}(\mathbf{x}) + \Delta I_\mathrm{n}(\mathbf{x})$, and the noise is characterized by its variance $\langle \Delta I_\mathrm{n}^2(\mathbf{x}) \rangle = \sigma_I^2(\mathbf{x})$, where $\sigma_I$ is the standard deviation of the noise in each voxel. 
This adds a term to the power spectrum estimator,
\begin{equation}
    \langle |\widetilde{\delta I}_\mathrm{s,n}(\mathbf{k})|^2 \rangle = \langle |\widetilde{\delta I}_\mathrm{s}(\mathbf{k})|^2 \rangle + \delta V \int \mathrm{d}^3\mathbf{x}\, w^2(\mathbf{x}) \sigma_I^2(\mathbf{x})\,.
\end{equation}
The second term still holds for larger than infinitesimal voxel volumes $\delta V$.

\subsection{Modeling the Sky Subtraction}
\label{subsec:sky_subtraction_modeling}
The sky subtraction is a common step in the reduction pipeline of ground-based optical data to remove zodiacal light, aurora, airglow, diffuse Galactic light, and emission from the Galactic warm interstellar medium \citep[WIM; e.g.,][]{wyse/gilmore:1992}. The optical sky spectrum consists of a continuum component and emission lines.
With the assumption that the sky spectrum is homogeneous on scales of the focal plane size, 
one can estimate the sky foreground by taking an average of the spectra in `empty' areas on the detector, 
i.e., areas that do not contain an object above a fiducial detection limit. 
The sky foreground spectrum is typically subtracted from the data. 

In addition to sky subtraction, further foreground mitigation is required. For example, continuum subtraction of the spectra can be applied to remove small-scale continuum foreground fluctuations, or wavelength regions around strong optical emission lines from the WIM of our Galaxy can be masked to reduce line foreground fluctuations. Masking spectral channels would add structure to the mask along the LOS, but can be modeled using the formalism in this work. Continuum foreground substraction removes the longest-mode fluctuations along the LOS, which can be reconstructed modeling suitable transfer functions \citep[see, e.g.,][]{cunnington/etal:2023b}{}{}. In this paper, we only model the sky subtraction for simplicity.

Sky subtraction also removes roughly the average intensity per redshift slice of the cosmological signal of interest
within the scale used for the sky foreground estimation, for example, the focal plane radius \citep[see, e.g., the discussion in][]{lujanniemeyer/etal:2022a}. Therefore, the sky subtraction decorrelates fluctuations on larger scales.
This does not affect the galaxy clustering of detected galaxies because the detection of galaxies is not influenced by the zeropoint on larger scales.
However, it does affect the intensity autopower and cross-power spectra even when all galaxies are detected because the intensity zeropoint is changed.
While optical ground-based LIM experiments such as HETDEX suffer from a loss of large-scale power due to sky subtraction,
ground-based LIM experiments in other wavelengths such as radio and submillimeter may also lose large-scale power due to similar issues.

We can model the effect of the sky subtraction on the intensity map by calculating the contribution of the line intensity to the estimated sky foreground.
We smooth the intensity map with a two-dimensional spherical top-hat kernel
in the plane perpendicular to the LOS with the size of the area used for the sky spectrum estimation.
Specifically, the estimated contribution to the sky foreground, i.e., the top-hat-smoothed intensity map, is
\begin{equation}
    \tilde{I}_\mathrm{sky} (\mathbf{k}) = \tilde{I}(\mathbf{k}) \tilde{D}_\mathrm{sky}(\mathbf{k}),
\end{equation}
where $\tilde{D}_\mathrm{sky}(\mathbf{k})$ is the Fourier transform of the two-dimensional spherical top-hat kernel, given by
\begin{equation}
    \tilde{D}_\mathrm{sky}(\mathbf{k}) = \frac{2 J_1\left(s_\mathrm{f} \sqrt{k_a^2+k_b^2} \right)}{s_\mathrm{f}\sqrt{k_a^2+k_b^2}} \,,
\end{equation}
where $a$ and $b$ denote the directions perpendicular to the LOS, and $J_1$ is the Bessel function of the first kind and of first order, and $s_\mathrm{f}$ is the radius of the area used for estimating the sky.

As $\tilde{D}_\mathrm{sky}(\mathbf{k})$ is real-valued, we obtain
\begin{equation}
    \langle |\tilde{I}(\mathbf{k}) - \tilde{I}_\mathrm{sky}(\mathbf{k})|^2 \rangle = \langle |\tilde{I}(\mathbf{k})|^2 \rangle \left[ 1 - 2 \tilde{D}_\mathrm{sky}(\mathbf{k}) + \tilde{D}_\mathrm{sky}^2(\mathbf{k})\right],\label{eq:pk_skysub_II_model}
\end{equation}
and 
\begin{equation}
    \langle \left[\tilde{I}(\mathbf{k}) - \tilde{I}_\mathrm{sky}(\mathbf{k})\right]\delta_\mathrm{g}^\ast \rangle = \langle \tilde{I}(\mathbf{k})\delta_\mathrm{g}^\ast \rangle\left[1 - \tilde{D}_\mathrm{sky}(\mathbf{k})\right].\label{eq:pk_skysub_Ig_model}
\end{equation}
This shows that the power spectrum is suppressed on scales larger than $s_\mathrm{f}$.

\section{Generating mock intensity maps}
\label{sec:simple_description}
This section describes the framework of the public code \textsc{Simple} for generating mock intensity maps.
In a nutshell, the code follows these steps:
\begin{enumerate}
    \item Generate a galaxy catalog in real and redshift space using the lognormal code of \citet{agrawal/etal:2017}. 
    \item Randomly assign line luminosities to galaxies following an input luminosity function.
    \item Assign redshift and flux to each galaxy. Apply the input selection function to distinguish between detected and undetected galaxies.
    \item Paint an intensity map and a galaxy density map using detected, undetected, or all galaxies in each map, and optionally smooth the intensity map.
    \item Optionally, generate and add an intensity noise map, apply a mask or weights, and model the sky subtraction.
    \item Calculate the auto- and cross-power spectra. 
\end{enumerate}

The main input parameters to \textsc{Simple} are cosmological parameters,
a luminosity function, a linear galaxy bias for all galaxies, 
the central redshift of the box, and the rest-frame wavelength or frequency of the target line.
If no tabulated matter power spectrum is provided as input, the cosmological parameters are used
to generate the matter power spectrum using the Eisenstein \& Hu fitting function \citep[][]{eisenstein/hu:1998}{}{}. 
They are also used to calculate the luminosity and angular diameter distances in later calculations.

Along with the luminosity function, one has to specify the minimum luminosity to obtain a finite number of galaxies.
This defines the number of galaxies to simulate with the lognormal galaxy catalog generator of \citet{agrawal/etal:2017}, 
which produces their positions, velocities, and redshift-space positions. 
This procedure assumes a flat sky and a single redshift.
We assign a luminosity to each galaxy by randomly drawing from the luminosity function. 
Unless otherwise specified, we define the first axis of the simulation box as the LOS and
assign the redshift to each galaxy according to its distance from the observer, inferred from the position in this simulation axis. 
We use this redshift to convert luminosities into fluxes.
Alternatively, a single redshift can be assigned to all galaxies.

We then apply a selection function based on an input flux limit above which a galaxy is detected, or on the target galaxy number density as a function of redshift.
This produces a galaxy catalog with detection flags.
We then calculate an intensity map and a galaxy number density map using the nearest-grid-point (NGP) assignment scheme.
One can specify which galaxies contribute to the intensity and galaxy maps, i.e., detected, undetected, or all galaxies. 

From the luminosities, we calculate the intensity map, 
either in terms of a specific intensity
\begin{equation}
\begin{split}
     I_\nu &= \frac{\mathrm{d}I}{\mathrm{d}\nu} = \frac{c \rho_L}{4 \pi H(z) \nu_0},\\
     \mathrm{or} ~
     I_\lambda &= \frac{\mathrm{d}I}{\mathrm{d}\lambda} = \frac{c \rho_L}{4 \pi H(z) (1+z)^2 \lambda_0},
\end{split}
\end{equation}
or brightness temperature
\begin{equation}
    T = \frac{\rho_L c^3 (1+z)^2}{8 \pi k_\mathrm{B} \nu_0^3 H(z)},
\end{equation}
where $\rho_L = \sum_{i=0}^{N_\mathrm{g}} L_i / \delta V$ is the total emissivity in each voxel, i.e., the sum of the luminosities of all galaxies in that voxel divided by the voxel volume, and $k_\mathrm{B}$ is the Boltzmann constant. This step assumes that the emission line is narrow, i.e. a delta function.

If specified in the input, the intensity map is smoothed with a Gaussian kernel perpendicular to the LOS, imitating the beam smoothing.
Along the LOS, one can apply Gaussian or top-hat smoothing, imitating a line-spread function or binning in redshift, wavelength, or frequency channels, respectively (see Section \ref{subsec:noise}).
The LOS smoothing can also be used to model broader lines than a voxel length.
One can add random Gaussian noise with the provided standard deviation per voxel $\sigma_{I}$, and subtract the estimated sky foreground (see Section \ref{subsec:sky_subtraction_modeling}) from the intensity map. If a mask is specified in the input, the galaxy number density and intensity maps are multiplied by the mask. 
The mask is equivalent to the weights introduced in Section~\ref{subsec:pk}.

Finally, one can calculate the summary statistics, i.e. the autopower spectra of galaxies and the intensity and the cross-power spectrum.
We compute the intensity and galaxy power spectra and the cross-power spectrum using the estimators defined in Section~\ref{subsec:pk}, where $V_{\rm box}$ is the volume of the simulation box.
We use the fast Fourier transform (FFT) to calculate $\widetilde{\delta I}$ and $\tilde{\delta}_\mathrm{g}$, keeping only the independent modes, and calculate the corresponding $\mathbf{k}$ and $\mu$ values of the cells. For the quadrupole, we multiply the mesh $\widetilde{\delta A}(\mathbf{k}) \widetilde{\delta B^\ast}(\mathbf{k})$ by the second Legendre polynomial evaluated at the $\mu$ values of the mesh. For each $k$ bin, we collect cells whose $k$ values fall into the bin and calculate their mean. We calculate the mean $k$ value of each bin in the same way.

\section{Example mock setup: HETDEX}
\label{sec:example_mock_setup}

In this section, we introduce the setup for a HETDEX-like \Lya LIM experiment. 
HETDEX is primarily a cosmological galaxy survey that aims to map approximately one million LAEs
through their \Lya emission line at $z\in [1.88,\,3.52]$ \citep{gebhardt/etal:2021}. 
HETDEX uses the Visible Integral-field Replicable Unit Spectrograph \citep[VIRUS;][]{hill/etal:2021} on the Hobby-Eberly Telescope (HET), 
which consists of $78$ integral-field unit spectrographs (IFUs), each of which contains $448$ fibers that are $\ang{;;1.5}$ in diameter, with a spectral resolution of $5.6$\,\AA.

\begin{table}
\centering
\begin{tabular}{|l|c|c|}
\hline
 & $\bar{z}=2.22$ & $\bar{z}=3.04$ \\ \hline
$L_\parallel^{\rm survey}$ {[}$h^{-1}$ Mpc{]} & 622 & 624 \\ \hline
$N_\mathrm{mock}$ & $7$ & $9$ \\\hline
$S_\perp^{\rm survey}$ {[}$h^{-2}$ Gpc$^2${]} & 2.38 & 3.23 \\ \hline
$f_V$ & 1.14 & 1.09 \\ \hline
$N_{\rm mesh}$ & 311 & 312 \\ \hline
$N_{\rm gal}/10^{6}$ & 14.2 & 14.4 \\ \hline
$\sigma_\parallel$ {[}$h^{-1}$ Mpc{]} & 1.76 & 1.27 \\ \hline
$s_f$ {[}$h^{-1}$ Mpc{]} & 10.0 & 11.6 \\ \hline
$\sigma_{I_\lambda}$ {[}$10^{-20}\mathrm{erg}\,\mathrm{s}^{-1}\,\mathrm{cm}^{-2}\,\mathrm{arcsec}^{-2}\,{\rm \AA}^{-1}${]} & 2.9 & 1.5 \\ \hline
\end{tabular}%
\caption{Summary of the Differences between the Low-$z$ and High-$z$ HETDEX Mocks.}
\label{tab:hetdex_mock_summary}
\end{table}

\begin{figure}
    \centering
    \includegraphics[width=0.48\textwidth]{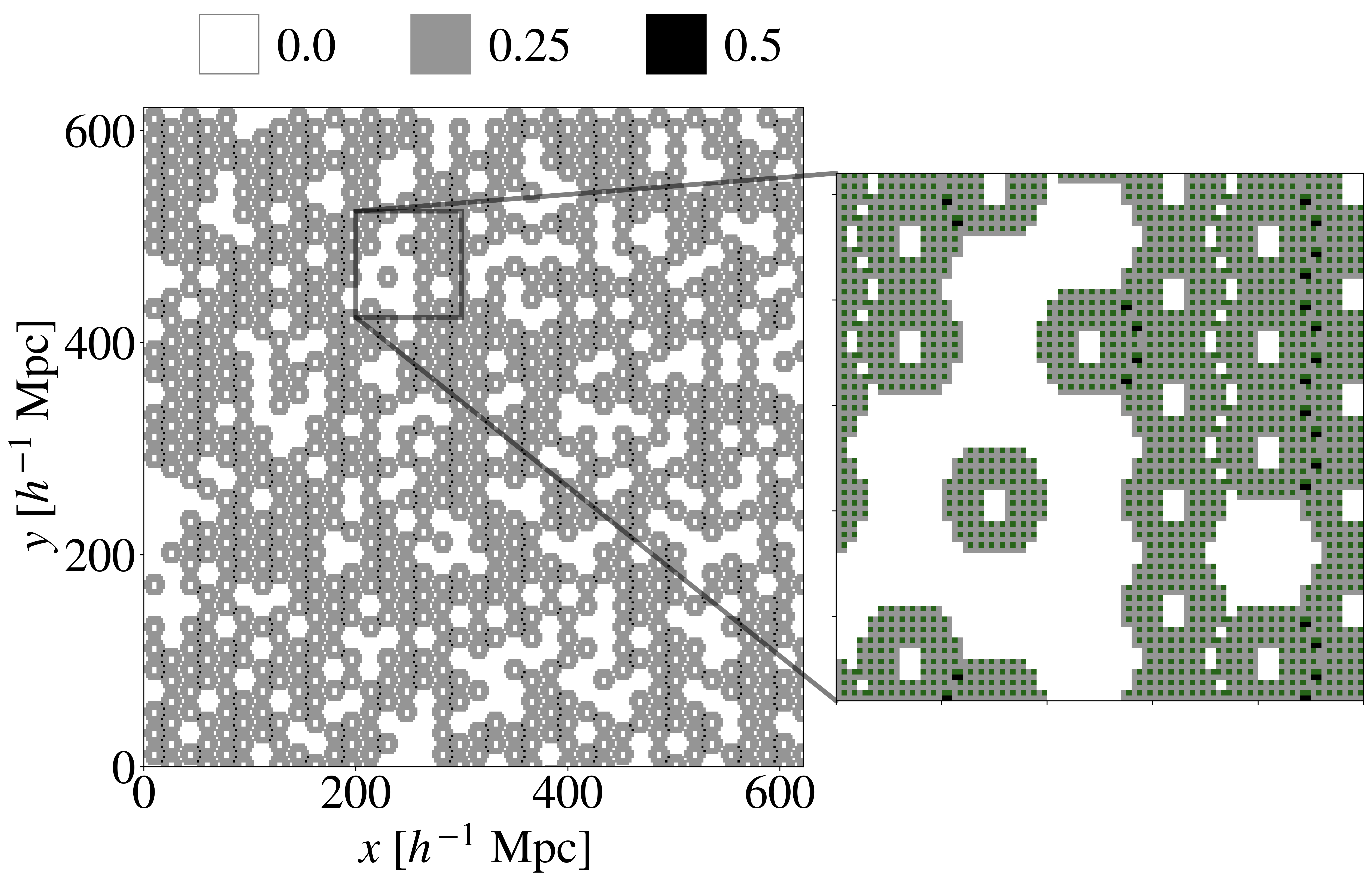}
    \caption{On-sky slice of a HETDEX-like mask applied to the HETDEX and validation mocks. The mask is constant in the LOS direction. The right panel is a zoom-in of a small area in the left panel. Each roughly hexagonal element with a hole in the center corresponds to one HETDEX observation with the VIRUS focal plane layout. The green squares in the right panel show the individual IFU positions in the higher-resolution mask that was used to generate the downsampled gray mask. The holes between observations show randomly removed individual observations to match the target fill factor of HETDEX.}
    \label{fig:hetdex_mask}
\end{figure}

HETDEX observes a total area of $540\,\mathrm{deg}^2$ without target preselection, expecting $460,000$ IFU observations. 
Because each IFU spans $\ang{;;51}\times\ang{;;51}$ in the sky, this amounts to an effective fill factor of the survey of $f_\mathrm{survey}\simeq 0.17$.
The layout of the IFUs across the $\ang{;18;}$ diameter focal plane leaves an IFU-sized gap between adjacent IFUs and a hole in the center of the focal plane, which is used for other instruments (see Figure \ref{fig:hetdex_mask}).
Three six-minute exposures per HETDEX observation fill in the gaps between fibers, but not between IFUs,
so that each individual HETDEX observation has a fill factor of $\simeq 1/4.6$.
The HETDEX data reduction pipeline offers a sky-subtraction mode that estimates the sky spectrum from all IFUs simultaneously,
i.e., roughly from a discontinuous circular area $\simeq \ang{;9;}$ in radius.

We divided the redshift range of HETDEX into two parts with a similar LOS distance of $622\,(624)\,h^{-1}\mathrm{Mpc}$
for the low-$z$ (high-$z$) sample.
The low-$z$ (high-$z$) sample covers $z\in [1.88,\, 2.57]$ ($z\in [2.57 ,\, 3.52]$) and is centered around $\bar{z}=2.22$ ($\bar{z}=3.04$).
At the mean redshift, the survey area of $540\,\mathrm{deg}^2$ translates into a comoving area of $2.38\,h^{-2}\mathrm{Gpc}^2$ ($3.23\,h^{-2}\mathrm{Gpc}^2$).
We generated $7000$ ($9000$) mocks of cubic volumes with side length $622\,h^{-1}\mathrm{Mpc}$ ($624\,h^{-1}\mathrm{Mpc}$).
By averaging the power spectra over $7$ ($9$), cubic boxes, we effectively obtained $1000$ power spectra of a box that is $1.14$ ($1.09$) times the size
of the low-$z$ (high-$z$) volume. In the last step, we multiplied the covariance matrix of each redshift slice by the respective factor to correct for this oversampling.
The largest accessible scale with this box size is $k_\mathrm{min} = 0.01\,h \mathrm{Mpc}^{-1}$.

We set the cosmological parameters to the fiducial cosmology (see Section \ref{sec:intro}). While the bias of detected LAEs is $\simeq 2$ \citep[][]{gawiser/etal:2007,guaita/etal:2010}{}{}, the bias of LAEs fainter than the detection limit is unknown. To obtain a conservative estimate for the signal-to-noise ratios (S/N) of the HETDEX power spectra, we choose a smaller linear galaxy bias of $b=1.5$.
We simulate the \Lya intensity at the rest-frame wavelength of $1215.67$\,\AA.
In the following, we use the specific intensity $I_\lambda$.
For both redshift sections, we adopt the \Lya luminosity function of the EWgt60 sample of \cite{konno/etal:2016} for galaxies at $z=2.2$, which is given by a Schechter function,
\begin{equation}
    \frac{{\rm d}n}{{\rm d}L} = \frac{\phi^\ast}{L^\ast} \left(\frac{L}{L^\ast}\right)^\alpha e^{-L/L^\ast}\,,
\end{equation}
with $L^\ast = 4.87 \times 10^{42}\,\mathrm{erg\, s^{-1}}$, $\phi^\ast = 3.37 \times 10^{-4}\,\mathrm{Mpc}^{-3}$, and $\alpha = -1.8$.
We set the minimum luminosity to $4\times 10^{40}\,\mathrm{erg}\,\mathrm{s}^{-1}$, based on Figure 4 of \citet{gronke/etal:2015}, and do not set the maximum luminosity.

\citet[][]{chiang/etal:2013} investigated the impact of sparse sampling for galaxy redshift surveys, in particular HETDEX, and found that the voxel size used for the power spectrum calculation has to be at least twice as large as the separation between IFUs to obtain an unbiased power spectrum.
For this reason, we grid the simulation box with a resolution of $2\,h^{-1}\mathrm{Mpc}$, i.e. $N_\mathrm{mesh}=311 \,(312)$, which corresponds to a Nyquist frequency of $k_\mathrm{Ny}=1.57\, h\mathrm{Mpc}^{-1}$.
Each voxel in our mock encompasses roughly two IFU side lengths ($\ang{;;102}$) perpendicular to the LOS and two spectral bins ($4$\,\AA) along the LOS, which is smaller than the spectral resolution.
This means that in our HETDEX-like experiment, one has to average the intensity of the fibers in each of the larger voxels. 

\begin{figure}
    \centering
    \includegraphics[width=0.46\textwidth]{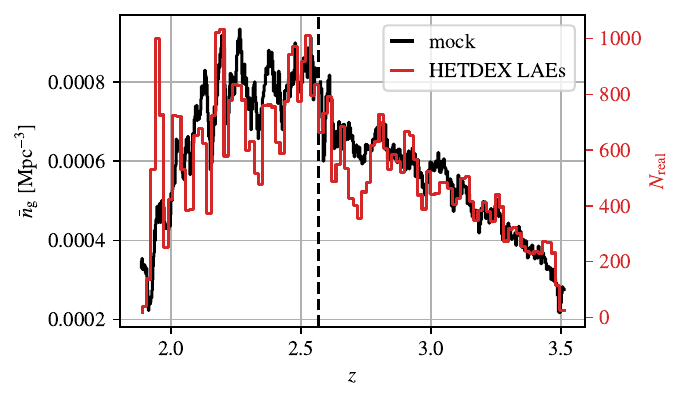}
    \caption{Mean detected galaxy number density as a function of redshift in one low-$z$ and one high-$z$ HETDEX-like mock (black). The dashed line shows the transition from the low-$z$ to the high-$z$ mock. The red histogram shows the distribution of detected LAEs in HETDEX with $\mathrm{S/N}>5.5$ from the HETDEX public source catalog I \citep[][]{mentuch-cooper/etal:2023}{}{}.}
    \label{fig:nofz}
\end{figure}

Each cubic box at low-$z$ (high-$z$) contains roughly $14.2$ ($14.4$) million galaxies, both detected and undetected.
We use a wavelength-dependent flux limit, enforcing
that there are $2.5$ detected galaxies per IFU on average, summed over the entire HETDEX LOS range \citep[see][]{gebhardt/etal:2021}.
This corresponds to a mean galaxy number density of $\bar{n}_\mathrm{g} \simeq 2.0 \times 10^{-3} \,h^3 \mathrm{Mpc}^{-3}$.
We rescaled the measured flux noise standard deviation from Figure 18 in \citet{hill/etal:2021} to obtain a wavelength-dependent flux limit that satisfies this condition. 

Figure \ref{fig:nofz} shows the resulting mean $\bar{n}_\mathrm{g}(z)$ in one realization of low-$z$ and high-$z$ HETDEX mocks.
We compare this to the distribution of LAEs from the HETDEX public source catalog I \citep[][]{mentuch-cooper/etal:2023}{}{}, where we selected LAEs with the `lae' flag in the `source\_type' column. The shapes agree well. A decrease in the detected LAEs at $z\simeq 2.7$ is caused by a mask applied at the center of $50\%$ of the HETDEX detectors and an increase in night-sky emission. 
The real number density in the high-$z$ volume can differ if the luminosity function at $z\simeq 3$ is different from that at $z\simeq 2$. However, because of the good agreement with the detected galaxies in HETDEX, we continue to use the luminosity function at $z=2.2$.

The high angular resolution of HETDEX allows one to mask individual detected galaxies without masking entire voxels. Therefore, we focus on the intensity map of only undetected sources. This intensity map does not share galaxies with the catalog of detected galaxies, so there is no contribution of shot noise to the cross-power spectrum between galaxies and intensity.\footnote{Combining the catalog of detected galaxies with the full intensity map of all sources adds information with respect to the autopower spectrum of detected galaxies alone. This is true even if the intensity map is dominated by bright detected galaxies (i.e., if the line luminosity function flattens at the faint end) because intensity fluctuations have a different (luminosity-weighted) bias from number-count fluctuations. We leave the study of this case for future work.}

We do not apply beam smoothing perpendicular to the LOS because the PSF of VIRUS is smaller than the size of an IFU/voxel. 
We smooth the intensity map along the LOS with a Gaussian kernel with standard deviation $\sigma=2.38$\,\AA ($\mathrm{FWHM}=5.6$\,\AA) to imitate the line-spread function, i.e., spectral resolution, of VIRUS \citep{hill/etal:2021}. This corresponds to $\sigma_\parallel=1.76\,h^{-1}\mathrm{Mpc}$
($1.27\,h^{-1}\mathrm{Mpc}$) for the low-$z$ (high-$z$) volumes. 
Because of the low spectral resolution and the large size of the voxels along the LOS, we do not model the broadening of the \Lya line. 
For the sky subtraction, we set the focal plane radius to $\ang{;9;}$, which corresponds to a distance scale of $s_\mathrm{f}=10.0\,h^{-1}\mathrm{Mpc}$ ($11.6\,h^{-1}\mathrm{Mpc}$) for the low-$z$ (high-$z$) part.

To add HETDEX-like noise, we transform the measured wavelength-dependent $5\sigma$ sky flux noise per resolution element per fiber in VIRUS \citep[see Fig. 18 in][]{hill/etal:2021}, $5\sigma_F$, into a specific intensity noise per fiber, $\sigma_{I_\lambda}$, by dividing by $\left(5\times \pi (\ang{;;0.75})^2 \times 5.6\,{\rm \AA} \right)$. Then we divide by $\sqrt{3\times 448}$ to account for the averaging over fibers within an IFU in three dithers.
The simulation voxels are shorter along the LOS than one spectral resolution element of VIRUS \citep[$5.6$\,\AA; see ][]{hill/etal:2021}{}{}. Because the noise below this scale is correlated, the factor of $\sqrt{\frac{5.6{\rm \AA}}{\Delta\lambda}}$ to obtain the noise in the voxel with LOS length $\Delta\lambda$ is an overestimate of the correlated noise. We therefore do not apply this factor. 
Because the area of a voxel encompasses four IFU areas, we divide this noise level by $\sqrt{N_\mathrm{IFU}(\mathbf{x})}$. Here, $N_\mathrm{IFU}(\mathbf{x})$ is the number of IFUs observed in the voxel at position $\mathbf{x}$, i.e. where the supersampled mask described below is nonzero. This can in principle be an integer between zero and four; for our masks, it is $\leq 2$.
In summary, we convert the $5\sigma$ flux noise per resolution element $5\sigma_F$ into the intensity noise at position $\mathbf{x}$ by calculating 
\begin{equation}
\sigma_{I_\lambda}(\mathbf{x}) = \frac{5\sigma_F}{5 \pi (\ang{;;0.75})^2 \sqrt{3\times 448\times N_\mathrm{IFU}(\mathbf{x})} \times 5.6\,{\rm \AA}}.
\end{equation}
This results in $\sigma_{I_\lambda}=2.9\,(1.5) \times 10^{-20} \,\sbunits$ in each voxel in the low-$z$ (high-$z$) boxes on average.

To obtain a mask on the coarse grid, we first generate a HETDEX-like mask with double resolution, so that each cell corresponds roughly to one IFU.
We generate a mask of VIRUS-like tiles of ones on and zeros in between IFUs and keep the mask constant along the LOS.
If we filled the entire area with observations, we would have a fill factor of $f_\mathrm{obs}=0.23$, 
similar to the focal plane fill factor of VIRUS $f^\mathrm{VIRUS}_\mathrm{obs}=1/4.6 = 0.22$. 
To match the effective fill factor of $f_\mathrm{survey} \simeq 0.17$ due to sparser observations, 
we randomly remove $\simeq 26\%$ of the individual observations after applying the VIRUS-like mask.
Then we downsample this mask to the same resolution as our simulated maps by averaging eight adjacent cells, which is equivalent to NGP assignment. The result is shown in Figure \ref{fig:hetdex_mask}. 
We generate $7$ ($9$) separate masks for the low-$z$ (high-$z$) boxes and then average the power spectra of the boxes with different masks. 

We calculate the power spectrum monopoles and quadrupoles in linearly spaced bins from $k_\mathrm{min} = 0.04\, h\mathrm{Mpc}^{-1}$ to $k_\mathrm{max} = 1\, h\mathrm{Mpc}^{-1}$ with $\Delta k =  0.04\, h\mathrm{Mpc}^{-1}$. We summarize the differences between the low-$z$ and high-$z$ HETDEX mocks in Table \ref{tab:hetdex_mock_summary}.

\section{Validation}
\label{sec:validation}

In this section, we show that the results of our simulations in real space agree with the expected power spectra given in Section \ref{sec:power_spectrum_modeling}. 
We perform this comparison in real space because
the lognormal algorithm precisely reproduces the input power spectra in real space, while the redshift-space power spectra deviate from the expectation in linear approximation \citep[see][]{agrawal/etal:2017}.

The setup of the validation mocks is almost identical to the low-$z$ HETDEX-like mocks described in Section \ref{sec:example_mock_setup}.
However, we reduce the intensity noise by a factor of $300$ and the flux limit by a factor of $5$ to reduce the shot noise for validation, and do not apply sky subtraction. 
We apply one of the HETDEX-like masks to all intensity maps.
We calculate the validation power spectra in real space and average over $1000$ mocks with side length $622\,h^{-1}\,\mathrm{Mpc}$.
All other input parameters are the same.
We calculate the intensity-intensity, galaxy-galaxy, and galaxy-intensity power spectrum monopoles and quadrupoles, where undetected galaxies contribute to the intensity map.

To obtain the model, we evaluate the input power spectrum that includes the galaxy bias on a mesh with the same $\mathbf{k}$ and $\mu$ values as obtained from FFT of the mock maps. We also evaluate the damping functions from the intensity smoothing on this mesh.
We calculate the window function by multiplying the mask by the mean expected intensity or the galaxy number density expected from the flux limit and the luminosity function as a function of redshift. We use FFT for the convolution of the power spectrum with the window function, add the shot noise, and multiply the result by the damping functions. Then we add the intensity noise power spectrum to the intensity autopower spectrum mesh.
We calculate the power spectrum multipoles of these results in the same way as for the mock (see Section \ref{sec:simple_description}).

\begin{figure*}
    \centering
    \includegraphics[width=0.8\textwidth]{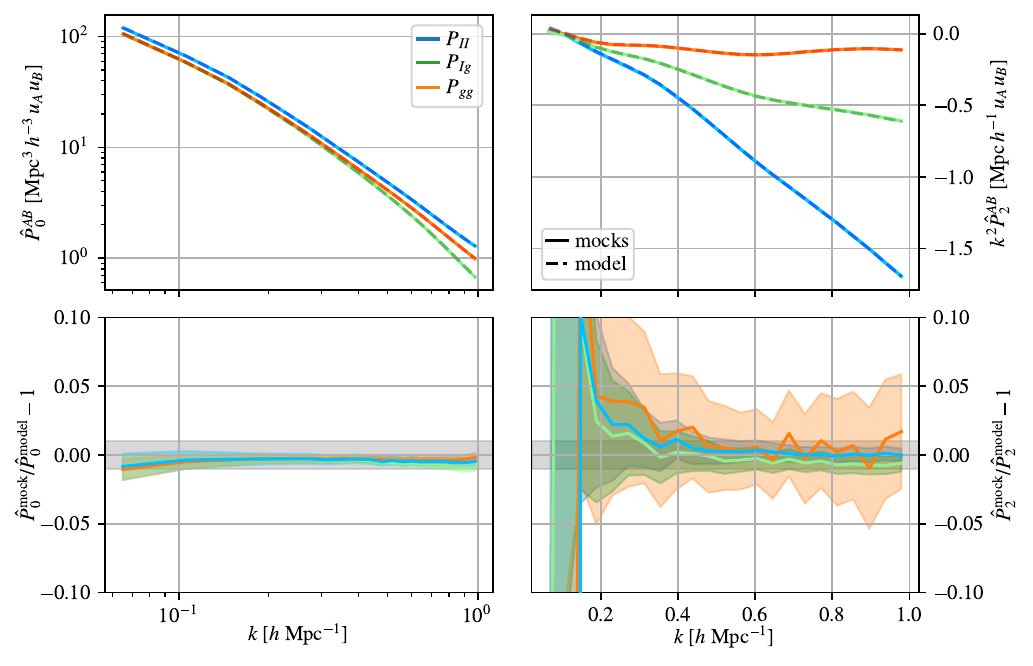}
    \caption{Comparison of the mock power spectra (solid lines) with the analytical model (dashed lines) of the monopole (left) and the quadrupole (right) power spectra in real space.
    The shot-noise (intensity noise) power spectrum is subtracted from the galaxy (intensity) autopower spectrum monopole.
    The bottom panels show the relative residuals between the mock and model power spectra. The units are $u_I = \langle I_{\lambda} \rangle \simeq 2.8\times 10^{-23}\,\sbunits$ and $u_\mathrm{g} = 1$. The shaded areas show the $1\sigma$ error of the mean given by the standard deviation of the different realizations divided by the square root of the number of realizations.
    The gray area shows deviations within $1\%$.}
    \label{fig:validation_realspace}
\end{figure*}

Figure \ref{fig:validation_realspace} shows the power spectrum monopole and the quadrupole in real space measured from the validation mocks. 
The shot-noise (intensity noise) power spectrum is subtracted from the galaxy (intensity) autopower spectrum monopole. 
The shot noise is not subtracted from the intensity autopower spectrum because it contains information about the luminosity function.
The quadrupole is affected by the anisotropic mask, the selection function, and the smoothing.
It also shows the analytical prediction presented in Section \ref{sec:power_spectrum_modeling}, as well as the relative residuals between the power spectra of the mock and the model.
The measured real-space power spectra from the mock agree with the model at all $k$ modes. 
We also tested the validation in the cases where all or only detected galaxies contribute to the intensity map and find excellent agreement.

\section{HETDEX forecast}
\label{sec:hetdex_forecast}
In this section, we forecast the sensitivity of HETDEX to LIM power spectra.
We use the mock setup described in Section \ref{sec:example_mock_setup} in redshift space and calculate the galaxy and intensity autopower spectra, as well as their cross-power spectrum with and without sky subtraction. As explained earlier, only undetected sources contribute to the intensity map.

\begin{figure*}
    \centering
    \includegraphics[width=0.8\textwidth]{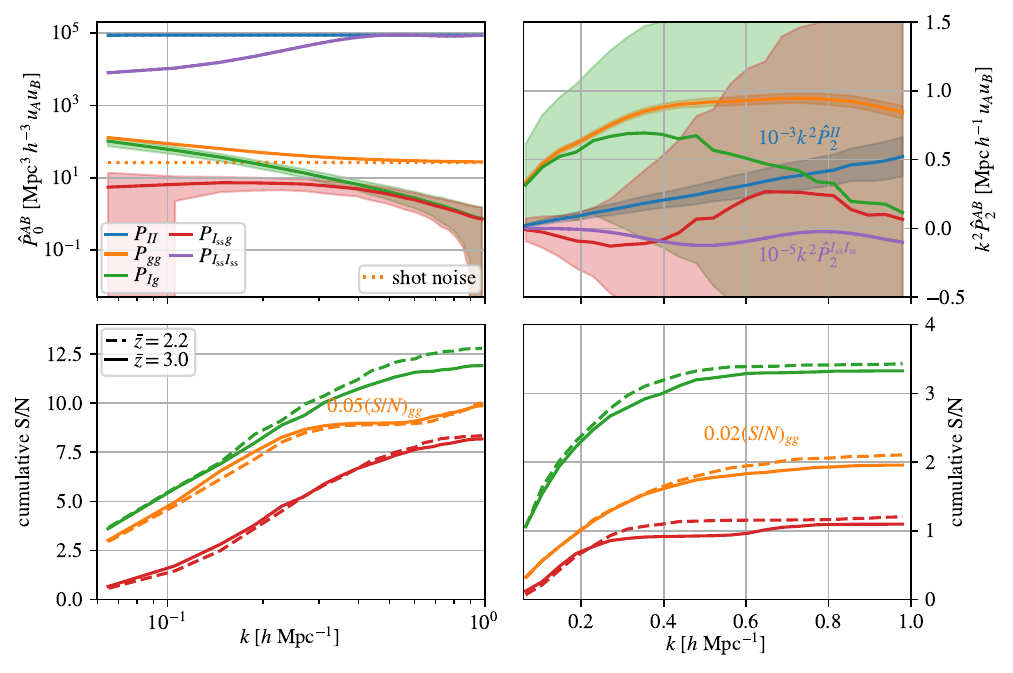}
    \caption{Power spectra of the HETDEX-like mocks in redshift space. The top left (right) panel shows the monopole (quadrupole) of the intensity (blue), galaxy (orange), and sky-subtracted intensity (purple) autopower spectra, and the intensity-galaxy cross-power spectra without (green) and with (red) sky subtraction
    in the high-$z$ volume.
    The quadrupole of the normal and sky-subtracted intensity autopower spectra were multiplied by factors of $10^{-3}$ and $10^{-5}$, respectively, for better visualization.
    The shaded areas show the square root of the diagonal elements of the corresponding covariance matrices.
    The units are $u_I = u_{I_\mathrm{ss}}=\langle I_\lambda \rangle \simeq 2.9 \times10^{-23}\,{\sbunits}$
    and $u_\mathrm{g}=1$.
    The dotted lines show the intensity noise (indistinguishable from the total power spectra) and the galaxy shot-noise power spectra.
    The bottom panels show the cumulative S/N up to a given $k$ bin after subtracting the shot noise from the galaxy autopower spectrum monopole and and quadrupole.
    The solid (dashed) lines correspond to the high-$z$ (low-$z$) HETDEX-like mocks.
    The cumulative S/N of the intensity autopower spectra are not shown because they are zero. The cumulative S/N of the galaxy autopower spectrum monopole (quadrupole) is multiplied by $0.05$ ($0.02$) for better visualization.}
    \label{fig:hetdex_forecast}
\end{figure*}

The upper panels of Figure \ref{fig:hetdex_forecast} show the monopole and quadrupole power spectra in redshift space.
The bottom panels show the cumulative S/N calculated as 
\begin{equation}
    \mathrm{S/N}_{AB,\ell}^2(k_N) = \bar{\mathbf{\Theta}}^T_{AB;\ell} \left(C^\mathbf{\Theta}_{AB,\ell}\right)^{-1} \bar{\mathbf{\Theta}}_{AB,\ell},
\end{equation}
where $k_N$ denotes the maximum wavenumber considered.
Here, $\mathbf{\bar\Theta}_{AB,\ell}=M^{-1}\sum_{i=1}^{M}\mathbf{\Theta}_{AB,\ell}^{(i)}$ is the mean of 
\begin{equation}
\mathbf{\Theta}^{(i)}_{AB,\ell} = \left(\hat{P}_{AB,\ell}^{(i)}(k_0), \hat{P}_{AB,\ell}^{(i)}(k_1),\dots,\hat{P}_{AB,\ell}^{(i)}(k_N)\right) - \mathbf{Q}
\end{equation} 
over $M$ realizations, where
$\hat{P}^{(i)}_{AB,\ell}$ is the estimator of the power spectrum monopole ($\ell=0$) or quadrupole ($\ell=2$) calculated from the $i$th realization of maps $A, B\in \{I,\delta_\mathrm{g}\}$. 

Here, $\mathbf{Q}$ is the constant shot noise of the galaxy autopower spectrum monopole,
the noise power spectrum for the intensity autopower spectrum monopole, and zero otherwise.
The noise power spectra of the intensity autopower spectra were calculated by generating $7000$ ($9000$) pure noise mocks in the low-$z$ (high-$z$) bin, performing the sky subtraction, calculating their monopole and quadrupole power spectra, and averaging them over the realizations.
For simplicity, we leave out the subscript ${AB,\ell}$ from now on.
We do not subtract the shot noise from the intensity autopower spectrum because it contains information about the luminosity function.
Because the intensity autopower spectra are indistinguishable from the intensity noise power spectra, the cumulative S/N are not shown in the figure.

$\left(C^\mathbf{\Theta}_{k_N}\right)^{-1}$ is the inverse of the covariance matrix, whose elements are defined as
\begin{equation}
\label{eq:cov}
    C^\mathbf{\Theta}_{mn} = \frac{f_V}{M-1}\sum_{i=1}^{M}\left(\mathbf{\Theta}^{(i)}_m - \bar{\mathbf{\Theta}}^{(i)}_m \right) \left(\mathbf{\Theta}^{(i)}_n - \bar{\mathbf{\Theta}}^{(i)}_n \right).
\end{equation}
Only elements of the covariance matrix up to the maximum $k_N$ are considered for matrix inversion to obtain $\left(C^\mathbf{\Theta}_{k_N}\right)^{-1}$.
As explained in Section \ref{sec:example_mock_setup}, we multiply the covariance matrix by the factor $f_V = 1.14 ~ (1.09)$ at low-$z$ (high-$z$) to correct for slightly larger simulated volumes than the HETDEX volume.

We predict that an ideal HETDEX-like experiment can detect the galaxy-intensity cross-power spectrum monopole at \lowzmonosignificance in each redshift bin of the survey despite the significant loss of large-scale power from the sky subtraction.
The quadrupole of the galaxy-intensity cross-power spectrum is not detectable (the S/N is $1$).
The intensity autopower spectrum also cannot be detected.

\begin{figure*}
    \centering
    \includegraphics[width=0.85\textwidth]{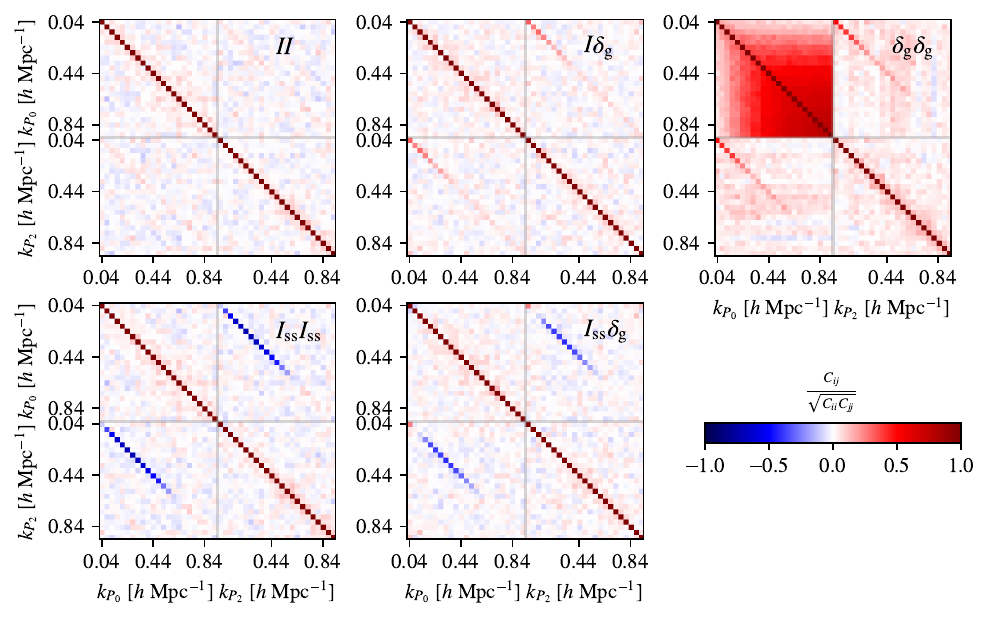}
    \caption{Correlation matrices of $\mathbf{\Phi}=\left(P_0(k_0), P_0(k_1),\dots,P_0(k_N),P_2(k_0),\dots,P_2(k_N) \right)$ for the intensity autopower spectrum, cross-power spectrum, galaxy autopower spectrum, sky-subtracted intensity autopower spectrum, and sky-subtracted cross-power spectrum (from top left to bottom right) of the high-$z$ HETDEX mock.}
    \label{fig:correlation_matrix_monoquadrupole_hetdex}
\end{figure*}

Figure \ref{fig:correlation_matrix_monoquadrupole_hetdex} shows the correlation matrices of the data vectors 
\begin{equation}
    \mathbf{\Phi}=\left(P_0(k_0), P_0(k_1),\dots,P_0(k_N),P_2(k_0),\dots,P_2(k_N) \right)
\end{equation} for the different power spectra that include the shot noise in the high-$z$ HETDEX mock.
The correlation matrix $R$ of a vector $\mathbf{\Phi}$ is given by
$R^\mathbf{\Phi}_{mn} = {C^\mathbf{\Phi}_{mn}}/\sqrt{C^\mathbf{\Phi}_{mm} C^\mathbf{\Phi}_{nn}}$,
where the covariance matrix is given in Eq.~\eqref{eq:cov}.

The galaxy autopower spectrum monopole has a large off-diagonal correlation that increases with increasing $k$. This is mainly due to the mode coupling introduced by the convolution with the complicated window function. The galaxy autopower spectrum monopole and quadrupole have a systematic low-level cross correlation that is visible as stripes at constant $k$ of the quadrupole. This may result from the imperfect integration of the $\mu$ values, in which the Legendre polynomials are no longer orthogonal \citep[see Appendix D.2.3 of][]{agrawal/etal:2017}{}{}.

The correlation matrices of the intensity autopower spectrum and the cross-power spectrum without sky subtraction are dominated by diagonal elements. They have an off-diagonal correlation between the monopole and quadrupole power spectra at the same $k$, which is slightly higher than the noise around most off-diagonal elements. Although this monopole-quadrupole correlation is positive without sky subtraction, it is larger and negative with sky subtraction.
The off-diagonal correlation for the intensity autopower spectrum monopole is negligible because the covariance is dominated by the uncorrelated intensity noise.
Because of the strong non-Gaussianity of lognormal realizations, the off-diagonal elements of the covariance matrix are overestimated \citep[][]{blot/etal:2019}{}{}. \textsc{Simple} therefore returns a conservative estimate of the S/N. This effect is most relevant for the galaxy power spectrum, the correlation matrix of which has the largest off-diagonal terms, as shown in Figure \ref{fig:correlation_matrix_monoquadrupole_hetdex}.

\section{Discussion}
\label{sec:discussion}

The \textsc{Simple} framework is a simple and fast
LIM simulation scheme with two main approximations.
First, the galaxy distribution is modeled by a lognormal random field following the input power spectrum that inculdes a linear bias \citep{agrawal/etal:2017}. The galaxies are obtained using Poisson sampling, and the galaxy velocities follow the linear continuity equation.
By construction, it is only accurate on the scales on which the input power spectrum is accurate, which, for example, is only true on linear scales for the Eisenstein \& Hu fitting function \citep[][]{eisenstein/hu:1998}{}{}. We may also underestimate the small-scale power spectrum because we do not model the one-halo term, nonlinear bias, or assembly bias.
It is possible to improve the density and velocity distribution of mocks at the expense of decreasing the speed of the code, for example by using other dark matter simulation methods described in the introduction.

The second approximation of the \textsc{Simple} framework is that the luminosity of a galaxy is randomly sampled from a luminosity function. Therefore, it does not depend on galaxy properties such as the star-formation rate or on its environment. This procedure misses any nontrivial connection between a specific galaxy population and their clustering, limiting any astrophysical analysis to the luminosity function. For instance, \textsc{Simple} can be used to study the performance of a given summary statistic to constrain the input luminosity function. Alternatives such as an empirical determination of the astrophysical properties, semi-analytic models of galaxy evolution, or hydrodynamical simulations defeat the purpose of \textsc{Simple}, given their computational cost.

\subsection{Limitations Specific to the \texorpdfstring{\Lya}{Ly-alpha} Line}
Our approach only models the emission from galaxies that is contained in the luminosity function. This does not capture all relevant physical emission processes of the \Lya line, as it neglects the recombination of ionized hydrogen and the collisional excitation of neutral hydrogen in the CGM and IGM \citep[e.g.,][]{dijkstra:2019}. Photons originating from a galaxy can also be observed far away from its source due to scattering in the CGM and IGM \citep{byrohl/etal:2021}{}{}.
\citet{byrohl/nelson:2023} find that while photons produced through diffuse emission in the IGM are negligible in the global \Lya luminosity budget, photons that originate from the ISM or CGM of galaxies and scatter in the IGM contribute substantially to the \Lya luminosity budget.
One can test the significance of the scattered contributions for LIM by modifying the luminosity function to include emission from the CGM calibrated on observations or by smoothing a small fraction of the intensity map to imitate scattering on larger scales than the voxel size. One can also add an intensity component that is directly proportional to the matter distribution to mimic diffuse gas emission.

Scattering of \Lya photons in the ISM strongly affects the escape fraction of \Lya photons and the emission peak wavelength \citep[e.g.,][]{hashimoto/etal:2013,blaizot/etal:2023}{}{}. The luminosity function contains the observed \Lya luminosity of detected galaxies, which consists only of photons that escaped from the ISM. However, the \textsc{Simple} framework does not account for the noncosmological redshift of the \Lya line due to radiative transfer within the ISM, which may cause an anisotropic effect similar to the Fingers-of-God effect \citep[][]{byrohl/saito/behrens:2019}{}{}.
It also does not account for radiative transfer in other environments, especially absorption, which may cause an anisotropic effect similar to the RSD, but with the opposite sign \citep[e.g.,][]{zheng/etal:2011,behrens/etal:2018}{}{}. These two effects can be modeled in an extension of the \textsc{Simple} code by shifting the \Lya line and calculating the optical depth along the LOS as a function of density and velocity, which can be obtained from a lognormal simulation.

\subsection{Limitations of the HETDEX Forecast}
This HETDEX forecast is meant as an order-of-magnitude forecast for the detectability of the cross-power spectrum of sky-subtracted intensity and detected galaxies. 
There are several possible improvements for a more accurate forecast in addition to the above \Lya modeling improvements:
one can account for the evolution of the luminosity function; include Galactic extinction; include interloper contamination of the intensity map, specifically [O\,II]-emitting galaxies at $z<0.5$, and masking of bright interlopers; model the mask, selection function, and non-Gaussian noise more accurately using the HETDEX data; account for false-positive galaxy detections; and model the intensity data reduction more accurately.
We leave these improvements for future work.

\section{Summary and conclusions}
\label{sec:summary}

We have presented the publicly available \textsc{Simple} code for quickly generating intensity maps that include observational effects such as noise, anisotropic smoothing, sky subtraction, and masking. 
It is based on a lognormal simulation of galaxies and random assignment of luminosities to these galaxies. Although this approach does not contain the dependence of the line luminosity on galaxy properties or of the galaxy properties on the cosmological environment, it provides a fast and versatile way to generate mock intensity maps. These can be used to study survey systematics and calculate covariance matrices of power spectra. 

We also derived an analytical model and showed that sky subtraction suppresses the power spectrum on scales larger than the focal plane size of the telescope.
We validated the SIMPLE code by showing that the power spectra in real space agree precisely with those of the analytical model.
This is the advantage of a lognormal mock, where the output power spectrum is designed to agree with the input power spectrum in real space \citep[][]{agrawal/etal:2017}{}{}.

As an application, we generated mock intensity and galaxy number density maps for a HETDEX-like LIM survey in redshift space that included a realistic mask, selection function, and intensity noise. We calculated the cross-power spectra of detected galaxies and the intensity of undetected galaxies after subtracting the sky spectrum and the respective covariance matrix in two redshift regimes within the survey. We predict that HETDEX will detect the cross-power spectrum monopole. 

In summary, \textsc{Simple} has been designed to provide fast but reliable realizations that allow for changes in cosmology, luminosity function, and observational specifications with little effort. For example, one can numerically estimate covariance matrices for different setups. Because lognormal realizations overestimate the off-diagonal elements of the covariance matrix due to strong non-Gaussianity, \textsc{Simple} returns a conservative estimate especially for the galaxy autopower spectrum \citep[][]{blot/etal:2019}{}{}. Therefore, this code is an excellent complement to slower, more physical simulations with the potential to guide the analysis of LIM surveys.

\begin{acknowledgments}
We thank S. Saito for helpful comments on the draft.
M.L.N. thanks L. Blot, M. Černetič, M. Fabricius, D. Farrow, G.J. Hill, D. Jeong, and J. Niemeyer for interesting and useful discussions.
J.L.B. acknowledges funding from the Ramón y Cajal Grant RYC2021-033191-I, financed by MCIN/AEI/10.13039/501100011033 and by
the European Union “NextGenerationEU”/PRTR.
\end{acknowledgments}

\vspace{5mm}
\software{astropy \citep{2013A&A...558A..33A,2018AJ....156..123A,astropy/etal:2022}; matplotlib \citep[][]{hunter:2007}{}{}; numpy \citep[][]{harris:2020}{}{}; scipy \citep[][]{2020SciPy-NMeth}{}{}; pmesh \citep[][]{yu_feng_2017_1051254}{}{}; dask \citep[][]{rocklin2015dask}{}{}; h5py \citep[][]{collette_python_hdf5_2014}{}{}}



\bibliography{sample631}{}
\bibliographystyle{aasjournal}

\end{document}